\documentclass[preprint,aps,amsmath,byrevtex,prd,titlepage,
nofootinbib]{revtex4-1}

\usepackage{dcolumn}
\usepackage{amsmath}
\usepackage{amssymb}
\usepackage{bm}
\usepackage{hyperref}
\usepackage{graphicx}
\usepackage{slashed}

\newcommand{\beq}{\begin{equation}}
\newcommand{\eeq}{\end{equation}}
\newcommand{\eq}[1]{Eq.~(\ref{#1})}

\begin{document}

\title {Light-by-Light Scattering Nonlogarithmic Corrections to Hyperfine Splitting in Muonium}
\author {Michael I. Eides}
%\altaffiliation[Also at ]{the Petersburg Nuclear Physics Institute,
%Gatchina, St.Petersburg 188300, Russia}
\email[Email address: ]{eides@pa.uky.edu, eides@thd.pnpi.spb.ru}
\affiliation{Department of Physics and Astronomy,
University of Kentucky, Lexington, KY 40506, USA\\
and Petersburg Nuclear Physics Institute,
Gatchina, St.Petersburg 188300, Russia}
\author{Valery A. Shelyuto}
\email[Email address: ]{shelyuto@vniim.ru}
\affiliation{D. I.  Mendeleyev Institute for Metrology,
St.Petersburg 190005, Russia}
%\date{}

\begin{abstract}
We consider three-loop corrections to hyperfine splitting in muonium generated by the gauge invariant set of diagrams with a virtual light-by-light scattering block. These diagrams produce both recoil and nonrecoil contributions to hyperfine splitting. Recoil corrections are enhanced by large logarithms of the muon-electron mass ratio. Both nonrecoil and logarithmically enhanced radiative-recoil corrections were calculated some time ago. Here we calculate nonlogarithmic radiative-recoil corrections generated by the insertions of the light-by-light scattering block.

\end{abstract}

%\pacs{12.20Ds,31.30.jf,32.10.Fn,36.10.Ee}
%\keywords{hyperfine splitting}

\preprint{UK/13-04}

\maketitle

\section{Introduction}

\begin{figure}[htb]
\includegraphics
[height=3cm]
{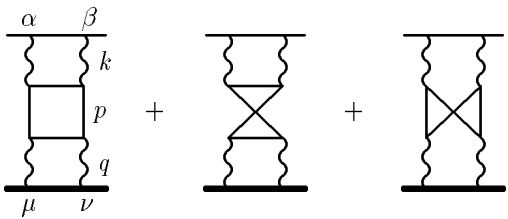}
\caption{\label{lblrec}
Diagrams with light-by-light scattering block}
\end{figure}

Theoretical and experimental research on hyperfine splitting (HFS) in the ground state of muonium has a long history, see e.g., \cite{egs2001,egs2007,mtn2012}. Measurement of the HFS in muonium is currently the best way to determine the value of the electron-muon mass ratio. Nowadays the HFS in the ground state of muonium is measured \cite{mbb,lbdd} with error bars in the ballpark of 16-51 Hz, and a new higher accuracy measurement is now planned at J-PARC, Japan \cite{shimomura}.
The value of $\alpha^2(m_\mu/m_e)$ is obtained from comparison of the HFS theory and experiment with the uncertainty that is dominated by the $2.3\times 10^{-8}$ relative uncertainty of the HFS theory \cite{mtn2012}. Improvement of the HFS theory would allow further reduction of the uncertainty of the electron-muon mass ratio. The current theoretical uncertainty of the HFS interval is  estimated to be about 70-100 Hz, respective relative error does not exceed $2.3\times10^{-8}$ (see discussions in \cite{egs2001,egs2007,mtn2012}). Reduction of the theoretical error of the HFS theory in muonium to about $10$ Hz is a  realistic goal \cite{egs2001,egs2007}.  Still unknown contributions include three-loop purely radiative corrections, three-loop radiative-recoil corrections, and nonlogarithmic recoil corrections (see detailed discussion in \cite{egs2007,mtn2012}) which are the main sources of the theoretical uncertainty. Below we consider three-loop radiative-recoil contributions to HFS generated by the light-by-light (LBL) scattering diagrams in Fig. \ref{lblrec} (and by three more diagrams with the crossed photon lines). These radiative-recoil corrections are additionally enhanced by the large logarithm of the electron-muon mass ratio. The logarithm squared and single-logarithmic terms are already calculated \cite{eks89,es2013}. Here we calculate the nonlogarithmic contribution.

We will follow the general approach to calculation of the three-loop radiative-recoil corrections to HFS developed in \cite{es0,eks89,egs01,egs03,egs04,esprd2009,esprl2009,esjetp2010} and start with the general expression for the LBL scattering contribution  in Fig. \ref{lblrec} (see, e.g., \cite{egs2001,egs2007})

\beq  \label{CROSS-1}
\Delta E =\frac{\alpha^2(Z\alpha)}{\pi^3}\frac{m}{M}E_F\biggl(-\frac{3M^2}{128}\biggr)
\int \frac{{d^4q}}{i\pi^2q^4}\biggl(\frac{1}{q^2+2Mq_0}+\frac{1}{q^2-2Mq_0}\biggl) T(q^2,q_0),
\eeq

\noindent
where

\beq  \label{T-def1}
\begin{split}
T(q^2,q_0)&=\frac12\int \frac{{d^4k}}{i\pi^2k^4}
\biggl(\frac{1}{k^2+2mk_0}+\frac{1}{k^2-2mk_0}\biggl) \langle\gamma^{\alpha}{\slashed k}\gamma^{\beta}\rangle
\langle\gamma^{\mu}{\slashed q}\gamma^{\nu}\rangle
S_{\alpha \beta \mu \nu}
\\
&=\langle\gamma^{\mu}{\slashed q}\gamma^{\nu}\rangle\int \frac{{d^4k}}{i\pi^2k^4}
\frac{ \langle\gamma^{\alpha}{\slashed k}\gamma^{\beta}\rangle}{k^2-2mk_0}S_{\alpha \beta \mu \nu},
\end{split}
\eeq

\noindent
$k^\mu$ is the four-momentum carried by the upper photon lines, $q^\mu$ is the four-momentum carried by the lower photon lines, $m$ is the electron mass, $M$ is the muon mass, $Z=1$ is the muon charge in terms of the electron charged used for classification of different contributions, and  $S_{\alpha\beta\mu\nu}$ is the light-by-light scattering tensor. The Fermi energy is defined as

\beq
E_F=\frac{8}{3}(Z\alpha)^4\frac{m}{M}\left(\frac{m_r}{m}\right)^3mc^2,
\eeq

\noindent
where $m_r$ is the reduced mass. The angle brackets in \eq{T-def1} denote the projection of the $\gamma$-matrix structures on the HFS interval (difference between the states with the total spin one and zero).

The integral in \eq{CROSS-1} contains both nonrecoil and recoil corrections to HFS that are partially already calculated (see \cite{egs2001,egs2007,es2013} for a collection of these results)

\beq \label{structure}
\begin{split}
\Delta E=
&\frac{\alpha^2(Z\alpha)}{\pi}(1+a_\mu)E_F[-0.472~514~(1)]
\\
&+\frac{\alpha^2(Z\alpha)}{\pi^3} E_F\frac{m}{M}
\left[\frac{9}{4}\ln^2{\frac{{M}}{m}}
+ \biggl(-3\zeta{(3)}- \frac{2\pi^2}{3} + \frac{91}{8}\biggr) \ln{\frac{{M}}{m}} +C_0\right],
\end{split}
\eeq

\noindent
where $a_\mu$ is the muon anomalous magnetic moment.

%\noindent
The leading nonrecoil term in \eq{structure} is generated by the nonrelativistic pole in the muon propagator

\beq  \label{NR-pole}
\frac{1}{q^2+2Mq_0+i0} \longrightarrow  -\frac{i\pi}{M}\delta (q_0),
\eeq

\noindent
and was calculated in \cite{eks1991,kn199496}. This is a numerically dominant contribution and it should be extracted analytically from the expression in \eq{CROSS-1} before calculation of the radiative-recoil corrections.

Recoil corrections generated by the diagrams in Fig. \ref{lblrec} contain three loop integrations and each of them could in principle generate a large logarithm of the electron-muon mass ratio. The strongly ordered region of integration momenta $m\ll k\ll p\ll q\ll M$  would produce logarithm cubed contribution but it turns into zero due to the tensor structure of the LBL block and fermion factors in this region \cite{es0}. The large logarithm squared, calculated in \cite{eks89}, arises from two integration regions, $m\ll k\sim p\ll q\ll M$ and $m\ll k\ll p\sim q\ll M$. Calculation of the single-logarithmic contributions is more involved and requires knowledge of the leading terms in the large momentum expansion of the function $T(q^2,q_0)$ in \eq{T-def1}. In \cite{es2013} after integration over the photon momenta $k$ and $q$ we obtained an integral representation for this function written as a sum of the ladder and crossed diagrams contributions in Fig.~\ref{lblrec}

\beq
T(q^2,q_0)=2T_L(q^2,q_0)+T_C(q^2,q_0).
\eeq

\noindent
The ladder contribution is represented as a sum of nine multidimensional integrals

\beq  \label{genint-L}
T_{L}(q^2, q_0)=\frac{128}{3}\int_0^1 {dy} \int_0^1 {dz}\int_0^1 {du}\int_0^1 {dt}
\sum_i {\cal T}_{L,i}(y, z, u, t, q^2, q_0),
\eeq

\noindent
where

\beq \label{calTL-1}
\begin{split}
{\cal T}_{L,1}& =  yz(1-t)(1-u)^2
\Biggr\{\biggl[\frac{1}{\Delta}-\frac{q^2 d^2}{\Delta^2} \biggr] (2q^2+q_0^2)
- \frac{ (q^2+2q_0^2)\tau^2}{\Delta^2}
- \frac{ q_0  (5q^2+q_0^2) \tau d}{\Delta^2} \Biggr\},
\end{split}
\eeq
\beq \label{calTL-2}
\begin{split}
{\cal T}_{L,2}& =\frac32 (2q^2+q_0^2)\Biggl\{-\frac{(1-2y) +2yz}{1-y}
\frac{(1-t)(1-u)^2}{\Delta}+(1-z) \frac{u(1-u)}{\Delta}
\\
&
-\frac{y^2z^2(1-z) q^2}{(1-y)^2}
\frac{(1-t) u(1-u)^2}{\Delta^2}\Biggr\},
\end{split}
\eeq
\beq \label{calTL-3}
\begin{split}
{\cal T}_{L,3}& =\Biggl\{\frac{(1-2y) +2yz}{1-y} \frac{(1-t)(1-u)^2}{\Delta^2}
-(1-z) \frac{u(1-u)}{\Delta^2}
\\
&+2 \frac{y^2z^2(1-z) q^2}{(1-y)^2}
\frac{(1-t) u(1-u)^2}{\Delta^3}\Biggr\} (2q^2+q_0^2) q^2d^2,
\end{split}
\eeq
\beq \label{calTL-4}
\begin{split}
{\cal T}_{L,4}& =
\Biggl\{\frac{(1-2y) +2yz}{1-y} \frac{(1-t)(1-u)^2}{\Delta^2}
\\
&-(1-z) \frac{u(1-u)}{\Delta^2}
+2 \frac{y^2z^2(1-z) q^2}{(1-y)^2}
\frac{(1-t) u(1-u)^2}{\Delta^3}\Biggr\}
\\
&\times\biggl[(2q^2+q_0^2)  \tau^2 + q_0 (5q^2+q_0^2) \tau  d\biggr],
\end{split}
\eeq
\beq \label{calTL-5}
\begin{split}
{\cal T}_{L,5}& =\frac{m^2}{1-y}\frac{(1-t)(1-u)^2}{\Delta^2}
\Bigl[(2q^2+q_0^2) d + 3 q_0\tau \Bigr],
\end{split}
\eeq
\beq \label{calTL-6}
\begin{split}
{\cal T}_{L,6}&  =4\int_0^1 {d\xi}\xi  yz^2 (1-t) u(1-u)^2
\\
&\times\Biggl\{\biggl[\frac34 \frac{1}{\Delta_{\xi}^2} - \frac{q^2d^2_{\xi}}{\Delta_{\xi}^3}\biggr]
(2q^2+q_0^2) q^2 d_{\xi}
- \frac{\tau^2  q_0^2 d_{\xi}}{\Delta_{\xi}^3}  (8q^2+q_0^2)
\\
&+\biggl[\frac14 \frac{1}{\Delta_{\xi}^2} - \frac{q^2d^2_{\xi}}{\Delta_{\xi}^3}\biggr]
(7q^2+2q_0^2) q_0\tau - \frac{3 q_0^3 \tau^3}{\Delta_{\xi}^3}\Biggr\},
\end{split}
\eeq
\beq \label{calTL-7}
\begin{split}
{\cal T}_{L,7}& =-\frac{yz(1-z)}{1-y}\frac{q^2 u(1-u)}{\Delta^2}
\Bigl[(2q^2+q_0^2) d + 3 q_0\tau \Bigr],
\end{split}
\eeq
\beq \label{calTL-8}
\begin{split}
{\cal T}_{L,8}& = 2 \frac{yz(1-z)}{1-y} (1-t) u(1-u)^2
\Biggl\{\biggl[-\frac34 \frac{1}{\Delta^2} + \frac{q^2d^2}{\Delta^3}\biggr]
(2q^2+q_0^2) q^2 d
\\
&+ \frac{\tau^2  q_0^2 d}{\Delta^3}  (8q^2+q_0^2)
+\biggl[-\frac14 \frac{1}{\Delta^2} + \frac{q^2d^2}{\Delta^3}\biggr]
(7q^2+2q_0^2) q_0\tau + \frac{3 q_0^3 \tau^3}{\Delta^3}\Biggr\},
\end{split}
\eeq
\beq
\label{calTL-9}
\begin{split}
{\cal T}_{L,9}&  =4 \frac{yz(1-z)}{1-y}(1-t) u(1-u)^2
\biggl[-\frac{1}{4} \frac{1}{\Delta^2} (2q^2+q_0^2) q^2d
\\
&+q^2(q^2-q_0^2) \frac{\tau^2 d}{\Delta^3}
- \frac{1}{4} \frac{1}{\Delta^2}(2q^2+q_0^2) q_0 \tau
+q_0(q^2-q_0^2) \frac{\tau^3}{\Delta^3}\biggr].
\end{split}
\eeq

\noindent
The crossed diagram contribution is represented as a sum of three multidimensional integrals

\beq  \label{genint-C}
T_{C}(q^2, q_0)=\frac{128}{3}\int_0^1 {dx}\int_0^1 {dy} \int_0^1 {dz}\int_0^1 {du}\int_0^1 {dt}
\sum_i {\cal T}_{C,i}(x, y, z, u, t, q^2, q_0),
\eeq

\noindent
where

\beq \label{calTC-1}
{\cal T}_{C,1} =\frac12 \frac{x(1-t)(1-u)^2}{1-xy}
\Biggr[(2q^2+q_0^2) \biggl[\frac{2}{\Delta}-\frac{q^2 d^2}{\Delta^2}\biggr]
- 3\frac{ q^2\tau^2}{\Delta^2}
- \frac{ q_0  (5q^2+q_0^2) \tau d}{\Delta^2} \Biggr],
\eeq
\beq  \label{calTC-2}
{\cal T}_{C,2} = \frac{x(1-t)(1-u)^2}{1-xy}
\frac{u m^2}{xy(1-xy)}
\biggl[\frac{2q^2+q_0^2}{\Delta^2}
-4 \frac{(q^2-q_0^2) \tau^2}{\Delta^3} \biggr],
\eeq
\beq
\label{calTC-3}
\begin{split}
{\cal T}_{C,3}& =\frac12 \frac{x(1-t)(1-u)^2}{1-xy}
 \Biggl[(2q^2+q_0^2) \frac{q^2 d^2}{\Delta^2}
-4(q^2-q_0^2) \frac{ q^2 \tau^2  d^2}{\Delta^3}
\\
&+(2q^2+q_0^2) \frac{q_0 \tau d}{\Delta^2}
-4(q^2-q_0^2) \frac{q_0 \tau^3  d}{\Delta^3}\Biggl].
\end{split}
\eeq

\noindent
In \eq{calTL-1}-\eq{calTC-3}

\beq  \label{Delta}
\begin{split}
&\Delta = g\Bigl[-q^2 + 2bq_0 + a^2\Bigr],
\quad
a^2=\frac{1}{g}\biggl[\tau^2+\frac{m^2u}{xy(1-xy)}\biggr],\quad
b=\frac{\tau d}{g},
\\
&d=\xi u\biggl[z- \frac{1-x}{1-xy}\biggr],\qquad \tau =m(1-u)t,
\qquad
g=g_0-d^2,
\\
&g_0=\frac{u(1-yz)(1-x+xyz)}{y(1-xy)},
\end{split}
\eeq

\noindent
and $x=1$ in \eq{calTL-1}-\eq{calTL-9}, while $\xi=1$ in all \eq{calTL-1}-\eq{calTC-3} except \eq{calTL-6}.

The large momentum expansions of the ladder and crossed functions $T_{L,C}(q^2,q_0)$

\beq  \label{ladder-asymp}
T_{L}\sim -\frac{16}{3}\frac{2q^2+q_0^2}{q^2}
\biggl[\ln{\frac{-q^2}{m^2}} - \frac{8\pi^2}{9}+ \frac{5}{6}\biggr]
-\frac{16}{3}\frac{q^2+2q_0^2}{q^2},
\eeq

\beq  \label{cross-asymp}
T_{C} \sim -\frac{64}{3} \frac{2q^2+q_0^2}{q^2}
\biggl[ \ln{\frac{-q^2}{m^2}} -2\zeta{(3)} + \frac{8}{3} \biggr]
- \frac{32}{3}\frac{q^2+2q_0^2}{q^2},
\eeq

\noindent
as well as the large momentum expansion of the total function $T$

\beq \label{total-asymp}
T=2T_L+T_C \sim -32\frac{2q^2+q_0^2}{q^2}
\biggl[\ln{\frac{-q^2}{m^2}}-\frac43\zeta{(3)} - \frac{8\pi^2}{27}+ \frac{37}{18}\biggr]-\frac{64}{3}\frac{q^2+2q_0^2}{q^2},
\eeq

\noindent
were calculated in \cite{es2013}. Both the already known double-logarithmic and the new single-logarithmic radiative-recoil contributions to HFS were obtained in \cite{es2013} from these large momentum expansions. Below we will use the exact explicit expressions for the function $T(q^2,q_0)$ to calculate a nonlogarithmic radiative-recoil contribution generated by the diagrams in Fig.~\ref{lblrec}.

\section{Calculation of Nonlogarithmic Contributions}

%\subsection{$\mu$- and $C$-Integrals}

In terms of the function $T(q^2,q_0)$ the total contribution to HFS of the diagrams in Fig.~\ref{lblrec} in \eq{CROSS-1} can be written in the form

\beq
\Delta E=\frac{\alpha^2(Z\alpha)}{\pi^3}E_F\frac{m}{M}J,
\eeq

\noindent
where

\beq \label{jint}
J=-\frac{3M^2}{128}\int \frac{d^4q}{i\pi^2 q^4}
\left(\frac{1}{q^2+2Mq_0}+\frac{1}{q^2-2Mq_0}\right)T(q^2,q_0).
\eeq

\noindent
We calculate this integral in Euclidean space and parameterize Euclidean  four-vectors $q_0=q\cos\theta$, $|\bm q|=q\sin\theta$. After the Wick rotation $\int d^4q\to (4\pi i/2)\int_0^\infty q^2dq^2\int_0^\pi d\theta\sin^2\theta$, and the integrand simplifies

\beq \label{vlumeel}
\frac{d^4q}{i\pi^2 q^2}
\left(\frac{M^2}{q^2+2Mq_0}+\frac{M^2}{q^2-2Mq_0}\right)
\to \frac{dq^2 d\theta\sin^2\theta}{\pi}
\frac{4M^2}{q^2+4M^2\cos^2\theta}.
\eeq

\noindent
Only the even in $q_0$ terms in the function $T(q^2,q_0)$ contribute to the integral in \eq{jint}. In order to simplify further integration we symmetrize the explicit expression for $T(q^2,q_0)$ with respect to $q_0$. All terms in \eq{calTL-1}-\eq{calTL-9} and \eq{calTC-1}-\eq{calTC-3} contain powers of the standard denominator $(-q^2 + 2bq_0 + a^2)$ (see definition in \eq{Delta}). It was shown in \cite{es2013} that one can neglect the term $2bq_0$ calculating the logarithmic contributions. Then after the Wick rotation it is convenient to write the symmetrized denominators inside the function $T(q^2,q_0)$ in the form

%\beq
\begin{align}
&\label{doloi_cos-1}
%\begin{split}
%&
\frac{1}{\left(-q^2+2bq_0+a^2\right)^n}
\longrightarrow \frac{1}{(q^2+a^2)^n} -{\cal E}_n \cos^2{\theta},
%\eeq
\\
%\\
%&
%\beq
\label{doloi_cos-2}
&\frac{q_0}{\left(-q^2+2bq_0+a^2\right)^n}
\longrightarrow {\cal O}_n \cos^2{\theta},
%\end{split}
%\eeq
\end{align}

\noindent
where

%\[
\beq
\begin{split}
{\cal E}_1& =\frac{4b^2q^2}{(q^2+a^2){\cal D}},
%\]
\\
%\[
{\cal E}_2 &= -\frac{\partial}{\partial a^2}{\cal E}_1
= \frac{4b^2q^2}{(q^2+a^2)^2{\cal D}^2}\Bigl[3(q^2+a^2)^2 + 4b^2q^2\cos^2{\theta}\Bigr],
%\]
\\
%\[
{\cal E}_3 &= \frac12 \biggl(\frac{\partial}{\partial a^2}\biggr)^2{\cal E}_1
%\]
%\[
= \frac{8b^2q^2}{(q^2+a^2)^3{\cal D}^3}\Bigl[3(q^2+a^2)^4 + 6(q^2+a^2)^2b^2q^2\cos^2{\theta}+ 8b^4q^4\cos^4{\theta}\Bigr],
%\]
\\
%\[
{\cal O}_1 &= \frac{2bq^2}{{\cal D}} ,
%\]
\\
%\[
{\cal O}_2 &= -\frac{\partial}{\partial a^2}{\cal O}_1
= \frac{4bq^2}{{\cal D}^2}(q^2+a^2) ,
%\]
\\
%\[
{\cal O}_3 &= \frac12 \biggl(\frac{\partial}{\partial a^2}\biggr)^2{\cal O}_1
= \frac{2bq^2}{{\cal D}^3}\Bigl[3(q^2+a^2)^2 - 4b^2q^2\cos^2{\theta}\Bigr],
%\]
\end{split}
\eeq

\noindent
and

\beq  \label{doloi_cos-3}
{\cal D} = (q^2+a^2)^2+4b^2q^2\cos^2{\theta}.
\eeq

\noindent
The numerators on the LHS in \eq{doloi_cos-1} and \eq{doloi_cos-2} can be multiplied by polynomials in $q^2$ and $q_0^2$.  These polynomials on the RHS turn into polynomials in $(-q^2)$ and $(-q^2\cos^2\theta)$.

The function $J$ in \eq{jint}  depends on $\mu=m/(2M)$ only due to the integrals containing in the integrand the first term on the RHS in \eq{doloi_cos-1}. We call these integrals $\mu$-integrals, and the general methods of their calculation are developed and described in \cite{eks1991ann,egs1998}. These $\mu$-integrals generate both nonrecoil and recoil contributions. Recoil contributions produced by the $\mu$-integrals contain logarithmically enhanced terms and $\mu$-independent contributions we are looking for.

The integrals of the other terms on the RHS in \eq{doloi_cos-1} and \eq{doloi_cos-2} ($C$-integrals) do not generate large logarithms and the corresponding recoil contributions  remain finite when $\mu$ goes to zero. Separate consideration of the $\mu$- and $C$-integrals  significantly simplifies further calculations.

The explicit expression for the integral $J$  in \eq{jint} after the Wick rotation has the form (we use the volume element in \eq{vlumeel})

\beq \label{wickene}
J=\frac{3}{128\pi}\int_0^\infty \frac{dq^2}{q^2}\int_0^\pi d\theta\sin^2\theta\frac{4M^2}{m^2q^2+4M^2\cos^2\theta}T(q^2,\cos^2\theta),
\eeq

\noindent
where we rescaled the integration momentum $q\to qm$. The function  $T(q^2,\cos^2\theta)$ is the same function as in \eq{jint} but with the Wick rotated momenta and after the substitutions  in \eq{doloi_cos-1} and \eq{doloi_cos-2}. As a result of rescaling this dimensionless function $T(q^2,\cos^2\theta)$ depends now on the dimensionless momentum $q$ and the parameter $m=1$ in \eq{calTL-1}-\eq{calTL-9} and \eq{calTC-1}-\eq{Delta}.

We are looking for the $\mu$-independent terms in the small $\mu$ (large $M$) expansion of the integral in \eq{wickene}. It is tempting to substitute $4M^2/(m^2q^2+4M^2\cos^2\theta)\to 1/\cos^2\theta$ directly inside the integrand in \eq{wickene}. Obviously this is not safe since the integral over $\theta$ can become divergent at $\cos\theta=0$ if an extra factor $\cos^2\theta$ is not supplied by the function $T(q^2,\cos^2\theta)$. Just by inspection we see that there are entries in the function $T(q^2,\cos^2\theta)$ that do not contain such a compensating factor. The reason for this spurious divergence at $\cos\theta=0$, or, what is the same, at $q_0=0$ is pretty obvious: $q_0=0$ corresponds to the nonrecoil contribution to HFS, and this spurious divergence is cutoff by $1/M$ in the original integral. This is the mechanism how an apparently recoil integral in \eq{wickene} produces a nonrecoil correction of order $1/\mu$. Hence, in case of such spurious divergence we cannot make the substitution $4M^2/(m^2q^2+4M^2\cos^2\theta)\to 1/\cos^2\theta$ inside the integral, and we need to calculate the integral over angles more accurately. By inspection we see that the integrals over angles in \eq{wickene} have the form

\beq \label{Aux-int-1}
\frac{4M^2}{\pi}\int_0^{\pi} {d \theta} \sin^2{\theta}
\frac{ \cos^{2n} {\theta}}{q^2 + 4M^2\cos^2{\theta}}
=\frac{1}{\pi}\int_0^{\pi} {d \theta} \sin^2{\theta}
\frac{ \cos^{2n} {\theta}}{\mu^2q^2 +\cos^2{\theta}}
=\Phi_n^{s}(q)+ \Phi_n^{\mu}(q),
\eeq

\noindent
where $n=0,1,2,3,\ldots$, and explicitly for $n=0,1,2,3$ (see \cite{egs1998})

\beq \label{Aux-int-1c}
\begin{split}
\Phi_n^s (q) & = \frac{\delta_{n0}}{\mu q},
\\
\Phi_0^{\mu}(q) &= \sqrt{1+\frac{1}{\mu^2 q^2}}- 1 - \frac{1}{\mu q},
\\
\Phi_1^{\mu}(q) &= -\mu^2 q^2\left(\sqrt{1+\frac{1}{\mu^2 q^2}}-1\right)+ \frac12,
\\
\Phi_2^{\mu}(q)&=\frac{1}{8} + \mu^2 q^2 \left[-\frac{1}{2} + \mu^2 q^2\left(\sqrt{1 + \frac{1}{\mu^2 q^2}}-1\right) \right],
\\
\Phi_3^{\mu}(q)&=\frac{1}{16} - \mu^2 q^2 \left\{\frac{1}{8} +
    \mu^2 q^2\left[-\frac{1}{2} + \mu^2q^2 \left(-1 + \sqrt{1 + \frac{1}{\mu^2 q^2}}\right) \right]\right\}.
\end{split}
\eeq

\noindent
Considering the integrand in \eq{Aux-int-1} and/or  the small $\mu$ expansions of the functions in \eq{Aux-int-1c}

\beq \label{expans}
\begin{split}
\Phi_0^{s}(q)+ \Phi_0^{\mu}(q)_{|\mu\to0}&\to \frac{1}{\mu q} - 1 + \frac{\mu q}{2},
\\
\Phi_1^{s}(q)+ \Phi_1^{\mu}(q)_{|\mu\to0}&\to \frac{1}{2} -\mu q,
\\
\Phi_2^{s}(q)+ \Phi_2^{\mu}(q)_{|\mu\to0}&\to \frac{1}{8} +O(\mu^2 q^2),
\\
\Phi_3^{s}(q)+ \Phi_3^{\mu}(q)_{|\mu\to0}&\to \frac{1}{16} +O(\mu^2 q^2),
\end{split}
\eeq

\noindent
we observe that only the integrals with $n=0$ generate singular at $\mu\to0$ contributions and do not admit the naive substitution $1/(\mu^2q^2+\cos^2\theta)\to1/\cos^2\theta$ in the integrand. Using the explicit expansions in \eq{expans} it is easy to check now that to separate the  nonrecoil ($1/\mu$) contributions in the integrals and simplify the calculation of $\mu$-independent terms in \eq{wickene} in the small $\mu$ case it is sufficient to make the substitution

\beq \label{prpresc}
\begin{split}
J&=\frac{3}{128\pi}\int_0^\infty \frac{dq^2}{q^2}\int_0^\pi d\theta\sin^2\theta\frac{1}{\mu^2q^2+\cos^2\theta}T(q^2,\cos^2\theta)
\\
&
\to
\frac{3}{128}\int_0^\infty \frac{dq^2}{q^2}\biggl[
\frac{1}{\mu q}T(q^2,\cos^2\theta=0)
\\
&+\frac{1}{\pi}\int_0^\pi d\theta\sin^2\theta %\wp
{\cal P}\left(\frac{1}{\cos^2\theta}\right)T(q^2,\cos^2\theta)\biggr].
\end{split}
\eeq

\noindent
Here we have introduced a new "principal value" prescription for integration over $\theta$

\beq
\frac{1}{\pi}\int_0^\pi d\theta
%\wp
{\cal P}\left(\frac{1}{\cos^2\theta}\right)=0.
\eeq

\noindent
As usual with the principal value $\wp({1}/{\cos^2\theta})\cos^2\theta=1$. Using this rule we can easily derive the rules for integration of the products $%\wp
{\cal P}({1}/{\cos^2\theta})$ with arbitrary polynomials of $\cos^2\theta$ and $\sin^2\theta$, for example

\beq
\frac{1}{\pi}\int_0^\pi d\theta\sin^2\theta
%\wp
{\cal P}\left(\frac{1}{\cos^2\theta}\right)=-1,
\qquad
\frac{1}{\pi}\int_0^\pi d\theta\sin^4\theta
%\wp
{\cal P}\left(\frac{1}{\cos^2\theta}\right)=-\frac{3}{2}.
\eeq

\noindent
These principal value prescriptions are justified  by the series expansions in  \eq{expans} for $n=0$ and by the explicit expression in the integrand in \eq{Aux-int-1} for any $n\geq1$.

The principal value prescription in \eq{prpresc} is a convenient and effective method for extracting the $\mu$-independent recoil corrections from the integral in \eq{wickene}. Still there remains a loophole. It was implicitly assumed that the integral over $q^2$ in the integral with the principal value in \eq{prpresc} is convergent at large momenta due to the function $T/q^2$, and effectively the integration momentum is bounded, $\mu q\ll1$. Clearly this assumption  is wrong for all terms generating logarithmically enhanced recoil corrections. Still, the leading logarithms arise exactly in the region $\mu q\ll1$ and we can use \eq{prpresc} to calculate these logarithms. We need to use the exact integrals in \eq{Aux-int-1} to calculate the nonleading logarithms and $\mu$-independent contributions in the case when $T/q^2$ does not guarantee convergence of the momentum integral in \eq{prpresc}.

After calculations we obtain nonlogarithmic contributions to HFS produced by the ladder

\beq \label{LADDER-TOTAL}
%\begin{split}
\Delta E_{L}=\frac{\alpha^2(Z\alpha)}{\pi^3} E_F \frac{m}{M}
[- 0.83071(5)],
%&= \frac{\alpha^2(Z\alpha)}{\pi} (1+a_\mu)E_F [ 0.40459\dots]
%\\
%&+\frac{\alpha^2(Z\alpha)}{\pi^3} E_F \frac{m}{M}  \Biggl[
%\frac{3}{8} \ln^2{\frac{M}{m}}+ \biggl(- \frac{\pi^2}{3} + %\frac{23}{16}\biggr) \ln{\frac{M}{m}}
%- 0.83071(5)\Biggr],
%\end{split}
\eeq

\noindent
and by the crossed diagrams

\beq \label{CROSS-TOTAL}
%\begin{split}
\Delta E_{C}=\frac{\alpha^2(Z\alpha)}{\pi^3} E_F \frac{m}{M}[7.65632(3) ].
%&=\frac{\alpha^2(Z\alpha)}{\pi}(1+a_\mu) E_F [-1.28156\dots ]
%\\
%&+\frac{\alpha^2(Z\alpha)}{\pi^3} E_F \frac{m}{M}  \Biggl[
%\frac32  \ln^2{\frac{M}{m}}
%+ \biggl(\frac{17}{2}-3\zeta{(3)}\biggr) \ln{\frac{M}{m}} + 7.65632(3) %\Biggr].
%\end{split}
\eeq

\noindent
The total nonlogarithmic recoil contribution to HFS generated by the diagrams in Fig. \ref{lblrec} is

\beq \label{TOTAL}
%\begin{split}
\Delta E^{nonlog}
%&
=2\Delta E_L+\Delta E_C
=\frac{\alpha^2(Z\alpha)}{\pi^3} E_F \frac{m}{M}[5.9949(1)]
\approx1.6~\mbox{Hz}.
%\frac{\alpha^2(Z\alpha)}{\pi}(1+a_\mu) E_F [ -0.472514(1)]
%\\
%&+\frac{\alpha^2(Z\alpha)}{\pi^3} E_F \frac{m}{M}
%\biggl[\frac{9}{4}  \ln^2{\frac{{M}}{m}}
%+ \biggl(-3\zeta{(3)} - \frac{2\pi^2}{3} + \frac{91}{8}\biggr) %\ln{\frac{{M}}{m}}
%+5.9949(1) \biggr].
%\end{split}
\eeq

\section{Conclusions}

Combining the new nonlogarithmic contribution to HFS in \eq{TOTAL} with the other contributions of the light by light scattering block in Fig. \ref{lblrec} calculated earlier \cite{egs2001,egs2007,es2013} we obtain the total contribution to HFS generated by these diagrams

\beq
\begin{split}
\Delta E=
&\frac{\alpha^2(Z\alpha)}{\pi}(1+a_\mu)E_F[-0.472~514~(1)]
\\
&+\frac{\alpha^2(Z\alpha)}{\pi^3}E_F\frac{m}{M}
\left[\frac{9}{4}\ln^2{\frac{{M}}{m}}
+ \biggl(-3\zeta{(3)}- \frac{2\pi^2}{3} + \frac{91}{8}\biggr) \ln{\frac{{M}}{m}} +5.9949(1)\right]
\approx-240.0~\mbox{Hz}.
\end{split}
\eeq

\noindent
This result makes us one step closer to calculation of all nonlogarithmic three-loop radiative-recoil corrections to HFS. Only two gauge invariant sets of diagrams with two radiative photon insertions either in the electron or the muon line remain uncalculated. We hope to report on the respective results in not so far future.

\begin{acknowledgments}

This work was supported by the NSF grant PHY-1066054. The work of V. S. was also supported in part by the RFBR grant 12-02-00313 and by the DFG grant GZ: HA 1457/7-2.

\end{acknowledgments}


\begin{thebibliography}{99}

\bibitem{egs2001} M.~I.~Eides, H.~Grotch, and V.~A.~Shelyuto, Phys. Rep.
\textbf{342}, 63 (2001).

\bibitem{egs2007} M.~I.~Eides, H.~Grotch, and V.~A.~Shelyuto, \textit{Theory
of Light Hydrogenic Bound States}, (Springer, Berlin, Heidelberg, New York, 2007).

\bibitem{mtn2012} P.~J.~Mohr, B.~N.~Taylor, and D.~B.~Newell, Rev. Mod. Phys. \textbf{84}, 1527 (2012).

\bibitem{mbb} F.~G.~Mariam, W.~Beer, P.~R.~Bolton et al, Phys. Rev.
Lett. \textbf{49}, 993 (1982).

\bibitem{lbdd} W.~Liu, M.~G.~Boshier, S.~Dhawan et al, Phys. Rev. Lett. \textbf{82}, 711 (1999).

\bibitem{shimomura} K.~Shimomura, AIP Conference Proceedings \textbf{1382}, 245 (2011).

\bibitem{eks89} M.~I.~Eides, S.~G.~Karshenboim, and V.~A.~Shelyuto,
Phys. Lett. B \textbf{216}, 405 (1989); Yad. Fiz. \textbf{49}, 493 (1989) [Sov. J. Nucl. Phys. \textbf{49}, 309 (1989)].

\bibitem{es2013} M.~I.~Eides and V.~A.~Shelyuto, Phys. Rev. D  \textbf{87}, 013005 (2013).

\bibitem{es0} M.~I.~Eides and V.~A.~Shelyuto, Phys. Lett. B \textbf{146},
241 (1984).

\bibitem{egs01} M.~I.~Eides, H.~Grotch, and V.~A.~Shelyuto, Phys. Rev.
D \textbf{65}, 013003 (2001).

\bibitem{egs03} M.~I.~Eides, H.~Grotch, and V.~A.~Shelyuto, Phys. Rev.
D \textbf{67}, 113003 (2003).

\bibitem{egs04} M.~I.~Eides, H.~Grotch, and V.~A.~Shelyuto,  Phys. Rev.
D \textbf{70}, 073005 (2004).

\bibitem{esprd2009} M.~I.~Eides and V.~A.~Shelyuto, Phys. Rev. D \textbf{80}, 053008 (2009).

\bibitem{esprl2009} M.~I.~Eides and V.~A.~Shelyuto, Phys. Rev. Lett. \textbf{103}, 133003 (2009).

\bibitem{esjetp2010} M.~I.~Eides and V.~A.~Shelyuto, JETP \textbf{110}, 17 (2010).

\bibitem{eks1991} M.~I.~Eides, S.~G.~Karshenboim, and V.~A.~Shelyuto,
Phys. Lett. B \textbf{268}, 433 (1991); \textbf{316}, 631 (E) (1993);
\textbf{319B}, 545 (E) (1993); Yad. Fiz. \textbf{55}, 466 (1992); \textbf{57}, 1343 (E) (1994) [Sov. J. Nucl.  Phys. \textbf{55},  257 (1992); \textbf{57}, 1275 (E) (1994)].

\bibitem{kn199496} T.~Kinoshita and M.~Nio, Phys. Rev.  Lett. \textbf{72}, 3803 (1994); Phys. Rev. D \textbf{53}, 4909 (1996).


\bibitem{eks1991ann} M.~I.~Eides, S.~G.~Karshenboim, and V.~A.~Shelyuto, Ann. Phys. \textbf{205}, 231, 291 (1991).

\bibitem{egs1998} M.~I.~Eides, H.~Grotch, and V.~A.~Shelyuto,  Phys. Rev.
D \textbf{58}, 013008 (1998).





\end{thebibliography}
\end{document}